
\input phyzzx

\REF\CJS{E. Cremmer, B. Julia and J. Scherk, Phys. Lett. {\bf 76B} (1978) 409.}
\REF\Ta{P.K. Townsend, Phys. Lett. {\bf B350}, (1995) 184.}
\REF\Wa{E. Witten, {\it String Theory Dynamics in Various Dimensions},
hep-th/9503124.}
\REF\BST{E. Bergshoeff, E. Sezgin and P.K. Townsend, Phys. Lett. {\bf 189B}
(1987) 75; Ann. Phys. (N.Y.) {\bf 185} (1988) 330.}
\REF\Bonan{E. Bonan, C.R. Acad. Sci. Paris, {\bf 262} (1962) 127.}
\REF\GPP{G.W. Gibbons, D. Page and C.N. Pope, Commun. Math. Phys. {\bf 127}
(1990) 529.}
\REF\KKS{M.J. Duff, B.E.W. Nilsson and C.N. Pope, Phys. Rep. {\bf 130} (1986)
1.}
\REF\DNP{M.J. Duff, B.E.W. Nilsson and C.N. Pope, Phys. Lett {\bf 129B}
(1983) 39; M.J. Duff and B.E.W. Nilsson, Phys. Lett. {\bf B175} (1986) 39.}
\REF\Tb{ P.K. Townsend, {\it String-Membrane duality in seven dimensions},
Phys. Lett. {\bf B}, {\it in press}.}
\REF\Gu {R. G{\" u}ven, Phys. Lett. {\bf 276B} (1992) 49.}
\REF\HTb{C.M. Hull and P.K. Townsend, {\it Enhanced gauge symmetries in
superstring theories}, Nucl. Phys. {\bf B} {\it in press}.}
\REF\DS {M.J. Duff and K.S. Stelle, Phys. Lett. {\bf 253B} (1991) 113.}
\REF\HTa{C.M. Hull and P.K. Townsend, Nucl. Phys. {\bf B438} (1995) 109.}
\REF\Sen{A. Sen, {\it String-String Duality conjecture in Six Dimensions and
Charged Solitonic Strings}, hep-th/9504027.}
\REF\HS{J. Harvey and A. Strominger, {\it The Heterotic String is a Soliton},
hep-th/9504047.}
\REF\WV{E. Witten and C. Vafa, {\it A one loop test of string duality},
hep-th/9505053.}
\REF\HT{P.S. Howe and P.K. Townsend, {\it Supermembranes and the modulus space
of $d=4$ superstrings}, in {\it Supermembranes and Physics in 2+1 dimensions},
eds.  M.J. Duff, C.N. Pope and E. Sezgin (World Scientific 1990).}
\REF\Sierra{G. Sierra, Phys. Lett. {\bf 157B} (1985) 379.}
\REF\BW{J. Bagger and E. Witten, Nucl. Phys. {\bf B222} (1983) 1.}
\REF\GST{M. G{\"u}naydin, G. Sierra and P.K. Townsend, Nucl. Phys. {\bf B242}
(1984) 244; Phys. Lett. {\bf 144B} (1984) 41; Phys. Rev. Lett. {\bf 53} (1984)
322; Nucl. Phys. {\bf B253} (1985) 573.}
\REF\KV{S. Kachru and C. Vafa, {\it Exact results for N=2 compactifications of
heterotic strings}, hep-th/9505105.}
\REF\FHSV{S. Ferrara, J.A. Harvey, A. Strominger and C. Vafa, {\it
Second-Quantized Mirror Symmetry}, hep-th/9505162.}
\REF\Narain{K. Narain, Phys. Lett. {\bf 169B} (1986) 41.}
\REF\NSW{K. Narain, M. Samadi and E. Witten, Nucl. Phys. {\bf B279} (1987)
369.}
\REF\Gin{P. Ginsparg, Phys. Rev. {\bf D35} (1987) 648.}
\REF\AGST{M. Awada, M. G{\" u}naydin, G. Sierra and P.K. Townsend, Class.
Quantum Grav. {\bf 2} (1985) 801.}
\REF\GNS{S.J. Gates, H. Nishino and E. Sezgin, Class. Quantum Grav. {\bf 3}
(1986) 21.}
\REF\CDLS{P. Candelas, A. Dale, C. L\"utken and R. Schimmrigk, Nucl. Phys. {\bf
B298} (1988) 493.}
\REF\GHL{P.S. Green, T. H\"ubsch and C.A. L\"utken, Class and Quantum Grav.
{\bf 6} (1989) 105.}
\REF\CLS{P. Candelas, M. Lynker and R. Schimmrigk, Nucl. Phys. {\bf B341}
(1990) 383.}
\REF\KS{A. Klemm and R. Schimmrigk, Nucl. Phys. {\bf B411} (1994) 559.}
\REF\Ja{D.D. Joyce, {\it Compact Riemannian 7-manifolds with holonomy $G_2$:1},
Oxford preprint.}
\REF\Jab{D.D. Joyce, {\it Compact Riemannian 7-manifolds with holonomy
$G_2$:2},
Oxford preprint.}
\REF\GSW{M.B. Green, J.H. Schwarz and E. Witten, {\it Superstring Theory: Vol
2}, (C.U.P. 1987)}
\REF\GSWest{M.B. Green, J.H. Schwarz and P.C. West, Nucl. Phys. {\bf B254}
(1985) 327.}
\REF\Strom{A. Strominger, {\it Massless black holes and conifolds in string
theory}, hep-th/9504090.}
\REF\Vafa{C. Vafa, {\it A stringy test of the fate of the conifold},
hep-th/9505023.}
\REF\Jb{D.D. Joyce, {\it Compact Riemannian 8-manifolds with holonomy Spin(7)},
Oxford preprint.}
\REF\CZV{M.T. Grisaru, A.E.M. van de Ven and D. Zanon, Phys. Lett. {\bf 137B}
(1986) 423; Nucl. Phys. {\bf B277} (1986) 388.}
\REF\HP{P.S. Howe and G. Papadopoulos, Phys. Lett. {\bf 263B} (1991) 230; {\it
ibid} {\bf 267B} (1991) 362; Commun. Math. Phys. {\bf 151} (1993) 467.}
\REF\HPW{P.S. Howe, G. Papadopoulos and P. West, Phys. Lett. {\bf 339B} (1994)
219.}
\REF\SV{S.L. Shatashvili and C. Vafa, {\it Superstrings and manifolds of
exceptional holonomy}, hep-th/9407025.}


\def\square{\kern1pt\vbox{\hrule height 0.5pt\hbox
{\vrule width 0.5pt\hskip 2pt
\vbox{\vskip 4pt}\hskip 2pt\vrule width 0.5pt}\hrule height
0.5pt}\kern1pt}


\Pubnum{ \vbox{ \hbox{R/95/31} \hbox{hep-th/9506150}} }
\pubtype{}
\date{June, 1995}

\titlepage

\title{Compactification of D=11 supergravity on spaces of exceptional
holonomy.}

\author{G. Papadopoulos and P. K. Townsend}
\address{DAMTP, University of Cambridge,
\break
Silver Street, Cambridge CB3 9EW,  U.K.}

\abstract {We investigate the compactification of D=11 supergravity to D=5,4,3,
on compact manifolds of holonomy $SU(3)$ (Calabi-Yau), $G_2$, and $Spin(7)$,
respectively, making use of examples of the latter two cases found recently by
Joyce. In each case the lower dimensional theory is a Maxwell/Einstein
supergravity theory. We find evidence for an equivalence, in certain cases,
with heterotic string compactifications from D=10 to D=5,4,3, on compact
manifolds of holonomy $SU(2)$ ($K_3\times S^1$), $SU(3)$, and $G_2$,
respectively.  Calabi-Yau manifolds with Hodge numbers $h_{1,1}=h_{1,2}=19$
play a significant role in the proposed equivalences.}

\endpage

\pagenumber=2

Supergravity theories exist in all spacetime dimensions $D$ with $D\le 11$. The
current wisdom is that, as quantum theories, they must be regarded as effective
field theories of superstring theories in some appropriate limit. Since the
critical dimension of superstring theories is $D=10$, the maximal $D=11$
supergravity [\CJS] has fallen into relative obscurity. Recently, however, it
has been realized that the $S^1$ compactified D=11 supergravity can be
interpreted as the effective field theory of the strongly coupled type IIA
superstring, with the Kaluza-Klein (KK) modes being identified with the quantum
states associated with the BPS-saturated extreme black holes of the $D=10$ IIA
supergravity theory [\Ta,\Wa]. It was further argued in [\Ta] that, for any
value of the string coupling, the type IIA superstring theory is, in
reality, an $S^1$ compactification of the D=11 supermembrane theory of [\BST].
The classical supermembrane exists in dimensions $D=11,7,5,4$ and (trivially)
in $D=3$, so in this context it is natural to consider compactifications of
$D=11$ supergravity to these dimensions. In each case one can then consider a
further compactification on $S^1$ to dimensions $D=6,4,3,2$, which (together
with D=10) are precisely those for which the classical Green-Schwarz
superstring action exists. Since we wish to preserve some supersymmetry, these
compactifications require consideration of compact spaces of (real) dimension
$4,6,7$ and $8$, with holonomy $SU(2)$, $SU(3)$, $G_2$, and $Spin(7)$,
respectively [\KKS]. All such spaces are Ricci flat (see [\Bonan,\GPP] for a
proof in the $G_2$
and $Spin(7)$ cases) so the lower dimensional field theory will have zero
cosmological constant. Flat tori of dimensions $4,6,7$ and $8$ are trivial
examples of such manifolds, because the holonomy group is trivial;
compactification on such spaces leads to a lower dimensional supergravity with
maximal supersymmetry. Here we shall be concerned with those
compactifications on spaces of dimension $4,6,7$ and $8$ for which the holonomy
is respectively $SU(2)$, $SU(3)$, $G_2$ and $Spin(7)$ but is not contained in
any proper subgroup. In this case the compactification preserves ${1\over2}$,
${1\over4}$, ${1\over8}$ and ${1\over 16}$ of the original supersymmetry,
respectively.

The only example in dimension $4$ is $K_3$. The compactification of D=11
supergravity on $K_3$ has been considered in [\DNP,\Wa,\Tb]. It leads to a D=7
Maxwell/Einstein (ME) supergravity theory with a three-form potential coupling
to the D=7 supermembrane. Evidence was given in [\Wa] that this ME supergravity
theory is the effective action for the $T^3$ compactified heterotic string at
strong coupling. In confirmation of this suggestion, it was further shown in
[\Tb] that the $T^3$ compactified heterotic string could be identified as the
solitonic magnetic fivebrane [\Gu] of D=11 supergravity, wrapped around the
$K_3$ compactifying space, and in [\HTb] that the symmetry enhancement known to
occur at special vacua of the weakly coupled $T^3$ compactified heterotic
string could be understood at strong coupling from consideration of the
wrapping modes of the electric membrane solution [\DS] of D=11 supergravity. We
note also that a duality between the $K_3\times T^2$ compactified type IIA
superstring and the $T^6$ compactified heterotic string was proposed in [\HTa],
and that evidence similar to that described above for D=7 has been found in
support of a D=6 duality relating the $K_3$ compactified type IIA string to the
$T^4$ compactified heterotic string [\Sen,\HS,\WV].

Manifolds of (real) dimension six with $SU(3)$ holonomy are called Calabi-Yau
(CY) manifolds, and many examples are known. The compactification of D=11
supergravity on CY manifolds was considered in [\HT]. It leads to D=5
supergravity coupled to $h_{1,1}-1$ vector multiplets and $h_{1,2}+1$ scalar
multiplets, where $h_{1,1}$ and $h_{1,2}$ are Hodge numbers. The scalar field
target space factorises [\Sierra] into a quaternionic sigma-model target space
[\BW] parameterized by the $4h_{1,2}+4$ scalars of the $h_{1,2}+1$ scalar
multiplets and a real manifold parametrised by the $(h_{1,1}-1)$ scalars of the
vector multiplets. The metric on the vector multiplet target space is
determined, as are all other couplings of the vector supermultiplet fields, by
a
homogeneous cubic polynomial in $h_{1,1}$ variables [\GST]. This polynomial is
determined by the Chern-Simons (CS) $FFA$ coupling of D=5 ME supergravity,
which has its origin in the similar CS term in D=11. To see this, we observe
that the harmonic expansion of the three-form potential $A$ of D=11
supergravity includes the term
$$
A= A^I\wedge \omega_I \qquad I= 1,\dots,h_{1,1}\ ,
\eqn\aone
$$
where $A^I$ are D=5 vector potentials and $\omega_I$ span the second cohomology
class of the CY manifold. Thus, the D=5 theory will include the term
$$
C_{IJK}\int_{M_5} F^I\wedge F^J\wedge A^K\ ,
\eqn\bone
$$
where $M_5$ is the five-dimensional spacetime and the numbers $C_{IJK}$ are
given by
$$
C_{IJK}= \int_{CY}\omega_I\wedge \omega_J\wedge\omega_K\ .
\eqn\cone
$$

In view of the relations just summarized between the $K_3$ compactification of
D=11 supergravity and the $T^3$ compactified heterotic string, it is tempting
to consider the possibility of similar relations between some
$CY$-compactification of D=11 supergravity (or a D=11 supermembrane theory) and
the heterotic string compactified on $K_3\times S^1$, as has been recently
explored in the D=4 context of CY compactifications of type II strings and
$K_3\times
T^2$ compactifications of the heterotic string [\KV,\FHSV]. To explore this
possibility we need to know the effective D=5 theory for the $K_3\times S^1$
compactified heterotic string. Consider first the compactification to D=9 on
$S^1$ of the heterotic string and choose the $S^1$ component of the $E_8\times
E_8$ or $SO(32)$ gauge potential such that the gauge group is broken to
$U(1)^{16}$. This yields the generic D=9 heterotic string theory, which has D=9
supergravity coupled to 17 vector multiplets as its effective field theory
[\Narain,\NSW,\Gin]. The bosonic field content of the D=9 graviton multiplet is
the graviton, a one-form potential, a two-form potential and a scalar. The
bosonic field content of the D=9 vector multiplet is a vector potential and a
scalar.  The  bosonic field content of D=9 supergravity coupled to 17 abelian
vector multiplets is therefore as follows: 1 graviton, 1 two-form potential, 18
abelian vector potentials and 18 scalars. The action was constructed
in [\AGST,\GNS]. We now compactify this D=9 theory on $K_3$ to obtain the
generic $K_3\times S^1$ compactification of the heterotic string. Taking into
account the fact that a two-form potential is equivalent in D=5, via duality,
to a vector potential, we find that the effective D=5 theory is N=2 (i.e.
minimal) D=5 supergravity coupled to 18 vector multiplets and 20
hypermultiplets. The sigma-model target space factorises, as mentioned above.
The target space of the vector multiplets can be deduced from the cubic
polynomial determined by the $FFA$ term in the D=5 action. The D=9 origin of
this term is the interaction [\AGST]
$$
\eta_{ab} \int_{M_9} A_5 \wedge F^a\wedge F^b \qquad (a=1,\dots,18)\ ,
\eqn\CSnine
$$
where $M_9$ is the D=9 spacetime, $A_5$ is the five-form gauge potential of the
dual action to the two-form version of D=9 supergravity, and $\eta_{ab}$
defines a quadratic form of Lorentzian signature in 18 variables. It follows
that the cubic polynomial in 19 variables of the D=5 theory must factorize into
a monomial and a quadratic polynomial in 18 variables of signature (17,1). It
then follows from the results of [\GST,\Sierra] that the target space for the
scalars in D=5 is
$$
\Big[SO(1,1)\times SO(17,1)/ SO(17)\Big] \times Q_{20}
\eqn\done
$$
where $Q_{20}$ is a quaternionic manifold of quaternionic dimension 20.

For this D=5 field theory to be obtainable by CY compactification of D=11
supergravity we need a CY manifold with Hodge numbers
$$
h_{1,1} = h_{1,2}= 19\ .
\eqn\one
$$
The Euler number of the CY space is therefore zero. A particular, complete
intersection, CY manifold with these Hodge numbers was constructed in [\CDLS]
and is tabulated as such in [\GHL]. Other CY manifolds with these Hodge numbers
can be found in [\CLS,\KS]. It is therefore possible that the compactification
of D=11 supergravity on one of the CY manifolds with these Hodge numbers will
lead to an effective D=5 action that can be identified with the effective
action of the $K_3\times S^1$ compactified heterotic string (at generic vacua),
but for this identification to be possible it is necessary that the cubic
polynomial determined by the intersection numbers $C_{IJK}$ of this CY manifold
factorise.

We now move on to 7-dimensional manifolds with $G_2$ holonomy. Examples of
compact manifolds with $G_2$ holonomy, and finite fundamental group, have been
constructed by Joyce [\Ja,\Jab]. The non-zero Betti numbers are $b_0=b_7=1$,
$b_2=b_5$ and $b_3=b_4$. We shall refer to these spaces as $J^{(7)}(b_2,b_3)$,
and we shall now determine the field content of the effective D=4 field theory
resulting from compactification of D=11 supergravity on these manifolds. The
bosonic field content of D=11 supergravity consists of the 11-metric and a
three-form gauge potential $A$. Consider first the 11-metric. This yields a
4-metric and $b_3$ scalars, $S^i,\, (i=1,\dots,b_3)$, because the number of
gauge-invariant perturbations of metrics of holonomy $G_2$ on a
seven-dimensional space is $b_3$ [\GPP]. Next, the three-form $A$ yields $b_2$
vectors, $A^I,\, (I=1,\dots,b_2)$, and $b_3$ pseudoscalars, $P^i$, in D=4 via
the ansatz
$$
A=
A^I\wedge \omega_I + P^i \Omega_i  \eqn\eone
$$
where the $\omega_I$ span the second cohomology group of $J^{(7)}(b_2,b_3)$ and
the $\Omega_i$ span the third cohomology group. Since the D=4 theory has N=1
supersymmetry, we already have sufficient information to determine its field
content, which must be N=1 supergravity coupled to $b_2$ N=1 vector multiplets
and $b_3$ scalar multiplets, each of which contains a scalar and a
pseudo-scalar field. It follows that the total number of D=4 Majorana spinor
fields (excluding the D=4 gravitino) that arise from the D=11 gravitino is
$b_2+b_3$.

Beyond the field content, we can also deduce certain features of the
interactions. Because of their D=11 origin, the pseudoscalars $P^i$ always
occur in interactions through their derivative $dP^i$, except in those
interactions that are due to the bare three-form potential $A$ in the D=11 CS
term, which leads to the D=4 interaction
$$
C_{IJ,k}\int_{M_4} P^k\, F^I\wedge F^J
\eqn\fone
$$
where $M_4$ is the four-dimensional spacetime, $F^I=dA^I$ are the two-form
field
strengths, and $C_{IJk}$ are the intersection numbers
$$
C_{IJ,k} = \int_{J_7} \omega_I \wedge \omega_J \wedge \Omega_k\ .
\eqn\gone
$$
Thus the pseudoscalars $P^i$ are axion fields. The target space metric for the
D=4 sigma model must take the form
$$
g_{ij}(S)dS^idS^j + h_{ij}(S) dP^idP^j\ .
\eqn\hone
$$
Moreover, this metric must be K\"ahler. Note that for such scalar field
interactions it is possible to replace the pseudoscalars $P^i$ by two-form
potentials. Thus another way of stating these restrictions on the target space
is to observe that the D=4 theory is N=1 supergravity coupled to $b_2$ vector
multiplets and $b_3$ `linear' multiplets.

One reason for interest in this compactification is a possible link between
D=11 supergravity and the much-studied CY compactified heterotic string. This
is because the effective action in the latter case is again an N=1 D=4
supergravity theory. The two compactifications cannot be equivalent in general
because the CY compactification of the heterotic string yields a {\it chiral}
D=4 theory with non-abelian gauge fields, whereas the $J^{(7)}$
compactifications of
D=11 supergravity yield non-chiral D=4 theories with abelian gauge fields.
However, if the CY manifold has zero Euler number then the resulting D=4
supergravity theory is non-chiral. Furthermore, since it is necessary to give a
non-zero expectation value to some $E_8\times E_8$ or $S0(32)$ field strengths,
$F$, on the CY manifold, in order to satisfy the condition that
[$tr(F^2)-tr(R^2)$] be cohomologous to zero, the non-abelian group is broken to
a subgroup, and may be further broken by Wilson lines (see, for example,
[\GSW]). We shall assume that the cohomology condition allows the breaking of
the gauge group to its maximal abelian subgroup by choosing $F$ to take values
in an abelian subgroup of $E_8\times E_8$ or $SO(32)$, as in [\GSWest].
The validity of this assumption depends on the topological properties of the
tangent bundle of the CY manifold.

This possibility motivates us to consider the compactification of the heterotic
string on CY manifolds with zero Euler number. We shall assume that the gauge
group of the heterotic string can be broken to $U(1)^{16}$ in the way just
suggested, and examine the consequences. Under these assumptions, the effective
D=4 theory is N=1 supergravity coupled to 16 N=1 vector multiplets and
$(h_{1,1}+h_{1,2}+1)$ N=1 scalar  multiplets. Because the Euler number
must vanish we further require that
$$
h_{1,1}=h_{1,2}\ .
\eqn\ione
$$
Thus, we have a D=4 supergravity coupled to 16 vector multiplets and
$(2h_{1,1}+1)$ scalar multiplets. Comparing with our previous results for the
$J^{(7)}$ compactifications of D=11 supergravity, we see that the field content
of the two compactifications agrees provided that
$$
b_2=16 \qquad b_3= 2h_{1,1}+1\ .
\eqn\jone
$$
Consider, for example, a CY manifold with $h_{1,1}=h_{1,2}=19$. In this case we
need
$$
b_2=16 \qquad b_3 =39\ .
\eqn\kone
$$
Remarkably, there is a compact seven-dimensional manifold with $G_2$ holonomy
having precisely these Betti numbers. It is the fourth example, for $l=8$, in
[\Jab].

There is a potential difficulty with the symmetry breaking mechanism outlined
above. Since the cohomology constraint has not been solved by the usual trick
of embedding the spin connection in the gauge group, the four-form
$tr(F^2)-tr(R^2)$ does not vanish identically, even though it is zero in
cohomology. It follows that the two-form gauge potential, $H$, of
the heterotic string must be non-zero, and this presents difficulties for the
preservation of supersymmetry in D=4. For this reason, it is not clear to us
that it really is possible to break the non-abelian gauge group to its maximal
abelian subgroup in the way described for compactifications to D=4 and, at the
same time, preserve supersymetry. If not, it is still possible that the
$J^{(7)}(16,39)$ compactified D=11 supergravity is equivalent to the
effective action of the heterotic string compactified on a CY manifold with
$h_{1,1}=h_{1,2}=19$ at singular points in the $J^{(7)}(16,39)$ moduli space,
if some such points are associated with an enhancement of the abelian gauge
symmetry to a non-abelian symmetry, as appears to occur both in the $K_3$
compactification of D=11 supergravity [\HTb] and in the CY compactification of
type II superstrings [\Strom,\Vafa]. In the example to hand, there are membrane
wrapping modes for each of the 16 homology two-cycles of $J^{(7)}(16,39)$, and
it is therefore possible that modes associated with a particular two-cycle
become massless as this cycle degenerates to zero area. Unlike the case of
$K_3$
compactification, however, this is not guaranteed by supersymmetry alone, so
renormalization effects may change the picture.

The uncertainties about the gauge group of the CY compactified heterotic string
can be simply resolved by a further compactification to D=3. As before, the
effective field theory can be deduced in this case by first compactifying to
D=9 on $S^1$. The generic D=9 theory for this compactification is as described
above, and this field theory can then be compactified on the CY manifold to
determine the generic D=3 effective field theory of the $CY\times S^1$
compactified heterotic string. Taking into account the equivalence in D=3
between vector potentials and scalars, this effective field theory is found to
be N=2 supergravity coupled to $(h_{1,1}+h_{1,2}+18)$ N=2 scalar multiplets.
This is to be compared with the N=2 D=3 supergravity coupled to $(b_2 +b_3 +1)$
scalar multiplets found by compactification of D=11 supergravity on
$J^{(7)}(b_2,b_3)\times S^1$. A necessary condition for their equivalence is
therefore that
$$
b_2 + b_3 = h_{1,1} + h_{1,2} + 17\ .
\eqn\two
$$
This condition is satisfied by several examples of CY and $J^{(7)}$ manifolds,
including $J^{(7)}(16,39)$ and the CY manifold with $h_{1,1}=h_{1,2}=19$. Note
that the RHS of \two\ is invariant under the mirror symmetry
$h_{1,1}\leftrightarrow h_{1,2}$.

We may also compare the N=2 D=3 supergravity theories obtained as just
described with those found by compactification from D=10 of either the type IIA
or the type IIB superstring on $J^{(7)}$, or rather of their D=10 effective
field theories. Not surprisingly, the D=3 field content of the $J^{(7)}$
compactified type IIA superstring is the same as that of $J^{(7)}\times S^1$
compactified D=11 supergravity. A surprise is that one finds the same field
content from the type IIB superstring. This would have been guaranteed in
advance if $J^{(7)}$ were to be the product of $S^1$ with a six-dimensional
manifold, because the IIA and IIB superstrings are known to be equivalent (as
are the respective supergravity theories) on compactification on $S^1$, but
$J^{(7)}$ is not a product. This `coincidence' leads us to conjecture that the
$J^{(7)}$
compactified type IIA and IIB theories are in fact equivalent.

Finally, we turn to 8-dimensional compact manifolds of holonomy $Spin(7)$.
Examples have been found recently by Joyce [\Jb]. They are all simply connected
and have vanishing first Betti number. We propose to refer to these examples as
$J^{(8)}(B_2;B_3;B_4^+,B_4^-)$, where $B_k$ is the $k$th Betti number. Note
that $B_4=B_4^+ +B_4^-$ since a basis of the closed but not exact four-forms is
provided by those the are either self-dual or anti-self dual. The dimension
of the moduli space of $Spin(7)$ structures is $B_4^-+1$ [\GPP,\Jb]. Using this
result it is straightforward to determine the bosonic field content of the
effective D=3 field theory. The fermionic field content then follows from N=1
supersymmetry. The result is D=3 N=1 supergravity coupled to $(B_2+B_3+B_4^-
+1)$ N=1 D=3 scalar multiplets. The N=1 D=3 supergravity theories obtained in
this way may be compared with those found by $J^{(7)}$ compactification of
the heterotic string. Again assuming that the gauge symmetry group of the
heterotic string is broken to its maximal abelian subgroup, we find that the
effective D=3 theory is N=1 D=3 supergravity coupled to ($b_2+b_3+17$) N=1 D=3
scalar multiplets, so equivalence of the two D=3 supergravity theories is
possible if
$$
B_2+B_3+B_4^- = b_2+b_3+16\ .
\eqn\three
$$
Again, the suggested equivalence seems more likely to hold after a further
$S^1$
compactification to D=2, since in this case the generic gauge symmetry of the
heterotic string is certainly abelian. The condition for equivalent field
content in D=2 is again \three. We remark that given any of the examples of
8-dimensional manifolds of $Spin(7)$ holonomy of [\Jb], there exists a
corresponding 7-manifold of holonomy $G_2$ among the examples given in [\Jab]
for which the condition \three\ is satisfied.

In D=2 we can also compare the above results with those of $J^{(8)}$
compactifications of type II string theories. The field content of the
effective D=2 theory obtained from the type IIA superstring is, not
surprisingly, the same as that found by $J^{(8)}\times S^1$ compactification of
D=11 supergravity; it is a locally supersymmetric sigma model with (1,1) D=2
supersymmetry. However, compactification of the type IIB superstring on
$J^{(8)}$ leads to very different results. The metric in D=2 has no kinetic
term, but imposes a constraint. For this reason it should be considered to
count as {\it minus} one degree of freedom. Taking this into account, the
bosonic field content of
the $J^{(8)}$ compactified type IIB superstring is $(2+ 2B_2 + B_4)$
left-moving scalars and $(2+ 2B_2 + B_4-\sigma)$ right-moving scalars, where
$\sigma$ is the signature $B_4^+-B_4^-$. Thus, the D=2 effective theory is {\it
chiral} if $\sigma\ne0$; in fact, it has (2,0) D=2 local supersymmetry. For all
of Joyce's examples, $\sigma =64$, so there is a chiral imbalance of 64 chiral
bosons. Since the D=10 type IIB theory is anomaly free, this effective chiral
D=2 theory must also be anomaly free. It would be of interest to determine the
fermion field content of this D=2 theory and verify the anomaly cancellation.

Note that these D=2 supergravity theories are not the worldsheet type of
supergravity theories of string theory. The latter have a role to play on
consideration of whether a given superstring compactification constitutes an
exact solution of the classical superstring theory. To be an exact solution,
the D=2 supersymmetric sigma-model having the compactifying space as its target
space must be conformally invariant. This condition is satisfied by $K_3$
because, being hyperKahler, the sigma model has (4,4) worldsheet supersymmetry.
In the case of sigma models with CY manifolds as target space, there is a four
loop divergence and the theory is not superconformally invariant [\CZV].
Supersymmetric sigma models with target spaces of holonomy $G_2$ or $Spin(7)$
were first studied in [\HP], where it was shown that the classical models are
invariant under $W$-symmetries associated with the covariantly constant forms
on their target spaces (as are the CY models). These sigma models are one-loop
finite (because they are Ricci flat) but they may have infinities at higher
loops. A free field realization of the W-current algebra introduced in [\HP]
has been constructed in [\HPW,\SV].

\vskip 1cm
\noindent{\bf Acknowledgements:} We are grateful to Rolf Schimmrigk for
information about Calabi-Yau manifolds. G.P. was supported by a Royal Society
University Research Fellowship.

\refout
\end